\documentclass[%
 reprint,
 superscriptaddress,
 amsmath,amssymb,
 prl,
 floatfix,
]{revtex4-1}

\usepackage{amsfonts,amsmath,latexsym,amssymb}
\usepackage{graphicx}% Include figure files
\usepackage{color}
\usepackage{dcolumn}% Align table columns on decimal point
\usepackage{bm}% bold math
\usepackage{hyperref}% add hypertext capabilities
\usepackage{enumerate}

\usepackage{graphics}
\usepackage{multirow}
\usepackage[caption=false]{subfig}
\usepackage{xcolor}

\begin{document}

%\preprint{APS/123-QED}

%%%%%%%%%%%%%%%%%%%%%%%%   MIRAR: una colección de posibles títulos    %%%%%%%%%%%%%%%%%%%%%%%%%%%%%%%%%%%%%

\title{The Polarized Monopole Liquid: a Coulomb phase in a fluid of magnetic charges}

%%%%%%%%%%%%%%%%%%%%%%%%%%%%%%%%%%%%%%%%%%%%%%%%%%%%%%%%%%%%%%%%%%%%%%%%%%%%%%%%%%%%%%%%%

\author{D. Slobinsky}
\affiliation{Departamento de Ingenier\'ia Mec\'anica, Facultad Regional La Plata, Universidad Tecnol\'ogica Nacional, Av. 60 Esq. 124, 1900 La Plata, Argentina}% \email{Second.Author@institution.edu}
\affiliation{{Instituto de F\'{\i}sica de L\'{\i}quidos y Sistemas Biol\'ogicos (IFLYSIB), UNLP-CONICET, La Plata, Argentina.}}

\author{L. Pili}
\affiliation{{Instituto de F\'{\i}sica de L\'{\i}quidos y Sistemas Biol\'ogicos (IFLYSIB), UNLP-CONICET, La Plata, Argentina.}}
\affiliation{Departamento de F\'\i{}sica, Facultad de Ciencias Exactas, Universidad 
Nacional de La Plata,  c.c.\ 16, suc.\ 4, 1900 La Plata, Argentina.}

\author{R. A. Borzi}
\affiliation{{Instituto de F\'{\i}sica de L\'{\i}quidos y Sistemas Biol\'ogicos (IFLYSIB), UNLP-CONICET, La Plata, Argentina.}}
\affiliation{Departamento de F\'\i{}sica, Facultad de Ciencias Exactas, Universidad 
Nacional de La Plata,  c.c.\ 16, suc.\ 4, 1900 La Plata, Argentina.}
%\ead{{borzi@fisica.unlp.edu.ar}}

\date{\today}% It is always \today, today,

\begin{abstract}

%%%%%%%%%%%%%%%%%%%%%%%%%%%%%%%%%%%%%%%%%%%%%%%%%%%%%%%%%%%%%%%%%%%%%%%%%%%%%%%%%%%%%%%%%
The forging of strong correlations on decreasing temperature can take place without the arousal of conventional order. If this happens, as in some geometrically frustrated magnets, disorder can be a phenomenon more interesting than order itself. A Coulomb phase, for example, has critical-like pair-spin correlations, leading to neutron scattering \textit{pinch points} and emergent electromagnetism. Here we present a new instance of disorder in an Ising pyrochlore lattice: the \textit{Polarized Monopole Liquid} (PML), a dense monopole fluid with pinch points in the \textit{magnetic charge}-pair correlations. It is a phase of ``monopole matter" never considered before which, in principle, can be stabilized in real materials using a magnetic field and uniaxial stress along the [100] direction. To explain how the monopole correlations arise, we show that the PML is a Coulomb phase in which spin fluctuations cannot be assigned either to monopoles or to internal magnetic moments, but necessarily comprehend both degrees of freedom. 
We develop a simple but nontrivial method to Helmholtz decompose the spin field into a divergenceless and a divergenceful part in magnetic charge disordered pyrochlores that shows the appearance of pinch points associated to the divergenceful component in places where Bragg peaks are observed for the ``all-in/all-out" antiferromagnet.
\end{abstract}


\maketitle

\vspace{2pc}

%\tableofcontents

%%%%%%%%%%%%%%%%%%%%%%%%%%%%%%%%%%%%%%%%%%%%%%%%%%%%%%%%%%%%%%%%%%%%%
%%%%%%%%%%%%%%%%%%%%%%%%   Motivation   %%%%%%%%%%%%%%%%%%%%%%%%%%%%%
%%%%%%%%%%%%%%%%%%%%%%%%%%%%%%%%%%%%%%%%%%%%%%%%%%%%%%%%%%%%%%%%%%%%%
%%%%%%%%%%%%%%%%%%%%%%%%%%%%%%%%%%%%%%%%%%%%%%%%%%%%%%%%%%%%%%%%%%
\emph{Introduction.} 
% ------ Opening: internal degrees of freedom of cooperative spins --------
When a fluid of magnetic ions crystallizes in a pyrochlore lattice the nuclear positions get frozen, but their magnetic moments or \emph{spins} $\textbf{S}$ can fluctuate down to very low temperatures \cite{Ramirez1994,Ramirez1999}. Something analogous happens when the magnetic charge-like quasiparticles or \textit{monopoles} inhabiting this structure~\cite{Castelnovo2008,Morris2009,jaubert2009nat} in turn crystallize in an array of plus and minus monopoles. Quite remarkably, in this state of \textit{monopole matter}~\cite{Castelnovo2008,Tch2011,Sazonov2012,Borzi2013,BBartlett2014,Guruciaga2014,jaubert2015p,Rau2016,Udagawa2016} both the static charge and the fluctuating moment of the monopoles are derived from a single degree of freedom. It is said that $\textbf{S}$ ``fragments" into a conservative static field $\textbf{S}_m$
% ------- Fragmentation ----------
 (which describes the crystal of monopoles) and a dipolar-like fluctuating field $\textbf{S}_d$ (accounting for the internal magnetic moment of the  monopoles), with $\textbf{S}_m + \textbf{S}_d = \textbf{S}$~\cite{BBartlett2014,Jaubert2015}\footnote{In the general case, in addition to the previous fields, a harmonic contribution $\textbf{S}_{har}$ can also be present \cite{Bramwell2017}.}. The divergenceless component $\textbf{S}_d$ on this Helmholtz decomposition corresponds then to a ``Coulomb phase"~\cite{Henley2010}. 
%
% ---------- Coulomb phase ----------
Notably in spin ices, whose low temperature magnetism is described by a divergence free condition, Coulomb phases have been observed through their characteristic signature in neutron scattering patterns: diffuse bowtie shaped scattering known as \textit{pinch points}~\cite{Moessner1998,Fennell2007,Fennell2009,Henley2010,Chang2010,Fennell2012,Petit2012}. The Helmholtz decomposition of the \emph{fragmented Coulomb spin liquid} phase (FCSL)~\cite{Jaubert2015}, however, provided a first example of a Coulomb phase coexisting with magnetic monopole order, a fact that has very recently found experimental counterparts~\cite{Petit2016,Lefrancois2017,Canals2016}.

% ---------- Results description --------
In this paper we describe a new phase of monopole matter and consider how it could be stabilized in a real material. We call it the \emph{Polarized Monopole Liquid} (PML), since it can be thought of as the Monopole Liquid (ML) 
~\cite{Slo2018} partially ordered by a magnetic field $\textbf{H}~//$ [100]. 
Like the ML (and different to the FCSL) the monopole positions fluctuate now as in a fluid.  Reciprocally, unlike its unpolarized cousin (and like the FCSL), 
there are strong pinch points in the diffuse neutron scattering structure factor of the PML phase. 
Quite remarkably, and unlike any previous case we know, we also find pinch points for \textit{monopole}-correlations, pointing to a very peculiar monopole distribution in the PML. We show that a single emergent gauge field is behind both type of correlations, in a phase where the fluctuations of magnetic charge and magnetic moments are tied together.

By using a simple method to Helmholtz decompose the spin field we evidence that even the divergenceful component of the field $\textbf{S}_m$ has associated pinch points in its neutron scattering pattern. While the ML and FCSL are extremely interesting phases, there is still no clue on how to thermodinamically stabilize them. Here we propose a family of systems in which a PML could be produced by the joint action of uniaxial stress and $\textbf{H}$.

    \begin{figure}
        \includegraphics[width=\columnwidth]{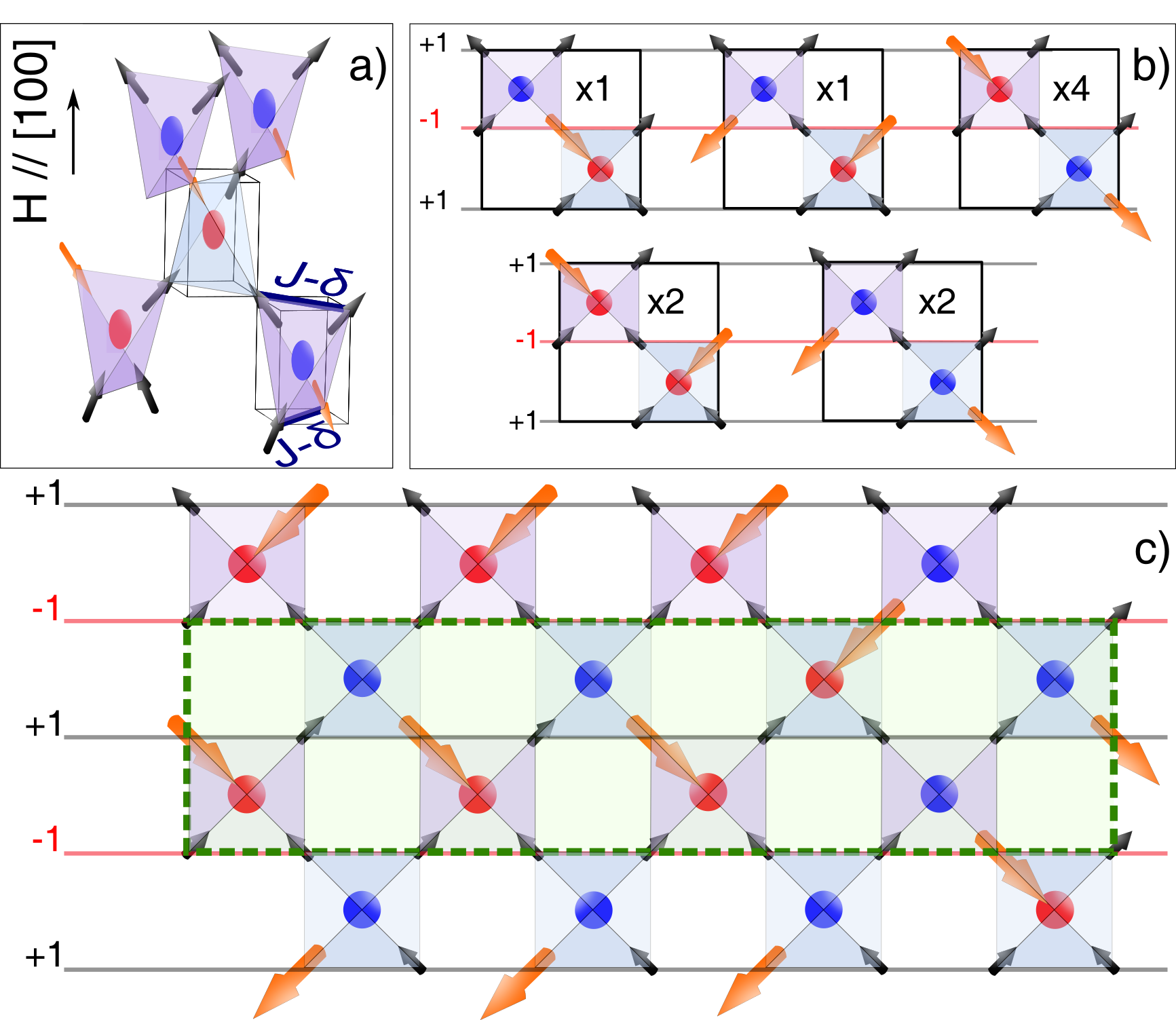}
        \caption{
        a) The pyrochlore lattice, with one ``up'' and four ``down'' tetrahedra. The vertical arrow indicates the field direction in all three panels, colored spheres stand for the two types of monopole-charge, and big orange arrows mark the ``dimer-spins" on a PML configuration. Two neighboring tetrahedra inscribed in cubes suggest how the checkerboard lattice can be thought of as a projection of the pyrochlore. The blue links in one cube indicate the relative change in the interaction constant due to stress along [100].
        b) The five possible monopole configurations of two contiguous PML tetrahedra conforming a \textit{block}, and their degeneracies. The spins in each [100] plane must be multiplied by $\pm1$ (as indicated by black and red lines) to obtain the divergenceless $\bm{\Sigma}$ and $\bm{\Sigma^{dim}}$ fields.
        c) Checkerboard lattice with an arbitrary PML configuration. An elongated Gaussian surface marked in green can be used to show that the total topological charge in alternate [100] planes is preserved.}
        \label{Fig1}
    \end{figure}
    %

%%%%%%%%%%%%%%%%%%%%%%%%%%%%%%%%%%%%%%%%%%%%%%%%%%%%%%%%%%%%%%%%%%%%%
%%%%%%%%%%%%%%%%%%%%%%%%%%%   Model   %%%%%%%%%%%%%%%%%%%%%%%%%%%%%%%
%%%%%%%%%%%%%%%%%%%%%%%%%%%%%%%%%%%%%%%%%%%%%%%%%%%%%%%%%%%%%%%%%%%%%
%%%%%%%%%%%%%%%%%%%%%%%%%%%%%%%%%%%%%%%%%%%%%%%%%%%%%%%%%%%%%%%%%%%%%
{\it The PML ground state.} 
% ----------- Defnition of the model -------------
Our base system is a pyrochlore array of Ising spins with $\langle 111 \rangle$ anisotropy in the vertices of ``up" tetrahedra (Fig.~\ref{Fig1}a);  panels b) and c) show the 
checkerboard lattice we use to simplify visualization when similar physics apply). We associate a positive single monopole to the center of a tetrahedron (a site of the ``dual" diamond lattice) when three of the spins point in and one out of it (a negative one for one-in--three-out). Its magnetic moment points along one of the four unit cell diagonals.
%
% ------------ Monopole liquid -------------------
In a ML each diamond site is occupied by either of the two magnetic charges, each having four different types of magnetic moments, with no further restrictions than those imposed by construction.

%
% ------------- PML from ML ---------------------
Since all spins have the same projection 
along [100], the PML ground state is equally populated by the four types of monopoles with a positive magnetic moment component along $\textbf{H}$, two of each sign. Each monopole can be uniquely labelled by its only spin with a negative component along [100], which always has its tail in a $-$ and its tip in a $+$ monopole below it (orange arrows in Fig.~\ref{Fig1}b) and c)). This spin (the \textit{dimer-spin}) can be identified with a hard-core dimer in the bonds of a diamond lattice. Conversely, any full coverage of this lattice by dimers maps into a unique PML spin configuration. 
The mapping permits to calculate the (finite) residual entropy per spin of this phase ($s / k_B = \frac{1}{4}\ln \left( \frac{27}{16} \right)\sim 0.13$~\cite{Nagle1966}), the same value as for the FCSL \cite{Jaubert2015}~\footnote{This case of a liquid and a crystal having similar entropies resembles that of $^3$He under pressure.}. We will see that, in contrast with the FCSL and with the ML~\cite{Slo2018}, this entropy is not associated with internal moments or monopole positions only, but with a combination of both.

In principle, the map to dimers would also allow to relate our system to the half magnetization plateau theory of CdCr$_2$O$_4$~\cite{Ueda2005}, which in its quantum version shows the appearance of artificial photons~\cite{Bergman2006b,Sikora2011}. The quantum PML is then a candidate to exhibit this kind of physics.

Fig.~\ref{Fig2}a) shows the monopole-monopole correlation function 
(see Sup. Mat.~\footnote{See Supplemental Material.}). Due to the constraints imposed by the applied field, the broad maxima around $\textbf{q}_{ch}=$ [111], [311], and [200] found in the ML~\cite{Slo2018} are now replaced by \emph{pinch points}. This reflects a unique situation within monopole matter systems which has not been anticipated: as with spins in Coulomb phases~\cite{Moessner1998,Henley2010} there can be long-ranged \textit{topological charge} correlations between $\textbf{q}_{ch}$ planes, with partial monopole disorder in each plane.

{\it The emergent gauge field.} 
% ----------- Average - fluctuation decomposition ------------
Since any spin $\textbf{S}$ will spend $1/4$ of its time pointing against $\textbf{H}$ it will have an ensemble average $\pm \textbf{S}/2$ (depending if it is pulled towards or against the center of its up tetrahedron). This finite average can be ascribed to a non-fluctuating harmonic field $\textbf{S}_{har}$. The corresponding Bragg peaks observed on the neutron scattering structure factor in the [$lhh$] plane are shown as black dots in Fig.~\ref{Fig2}b). In marked contrast with the ML~\cite{Slo2018}, the diffuse pattern related to $\Delta\textbf{S}=\textbf{S}-\textbf{S}_{har}$ yields pinch points. They sit in multiples of [200] and [111] (like spin ices in zero field \cite{Bramwell2001,Fennell2009}), with \emph{extra ones} in [311], [133], [022], and [422] (where the ``all-in/all-out" antiferromagnet, AIAO, shows Bragg peaks). 
%

% -------------- Charge correlations ---------------
 \begin{figure*}
      \centering
     % \subfloat[lal lal]
        {\includegraphics[width=\textwidth]{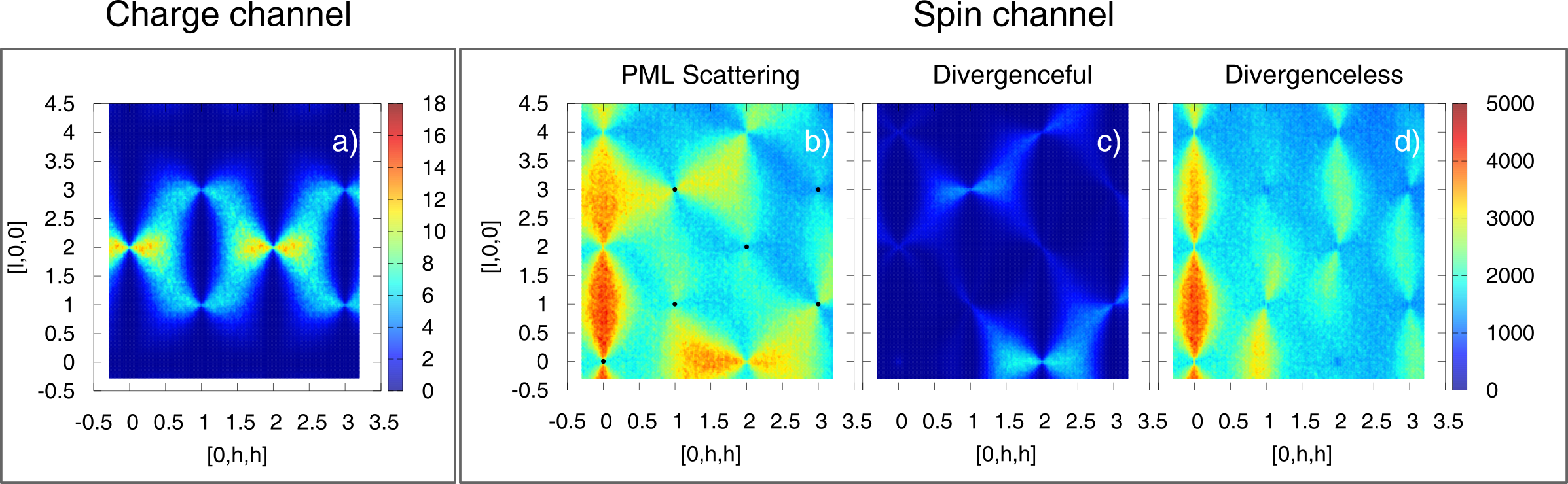}} 
      \caption{a) Monopole-monopole correlation in the $[lhh]$ plane for the PML, showing pinch points in the monopole channel.
      b) Diffuse pattern for the lattice spin field $\textbf{S}$, with Bragg peaks marked with black dots. The diffuse scattering presents pinch points around the same $\textbf{q}$ values as spin ice, but also where the AIAO antiferromagnetic phase shows Bragg peaks \cite{Slo2018}.
      c) Diffuse pattern for the approximated divergenceful component $\bm{S}_m$, showing intense pinch-points for $\textbf{q}=$[311] and [022].
      d) Diffuse pattern for the approximated divergenceless component $\bm{S}_d$, showing intense pinch-points along $\textbf{q}=$[200] and [111] directions. We have allowed for a maximum absolute charge within this channel of $10^{-4}$. 
      All colored plots were obtained using $~5\times10^5$ spins in $330$ independent configurations (more details in the Sup. Mat.).
      }
    \label{Fig2}
 \end{figure*}

%    
% --------------- Coulomb phases ---------------
Due to the correspondence we found between spin configurations and classical hard-core dimers on a bipartite lattice~\cite{Huse2003}, a Coulomb phase was not unexpected in the PML.
In fact, we can relate $\textbf{S}$ with a new (divergenceless) spin field $\bm{\Sigma}$, after redefining the vertices over which the net flow is evaluated.
We illustrate our procedure in two steps. 
$i-$  We establish a minimum volume or \textit{block} to be considered;
it comprises \textit{two} neighboring diamond sites ({\it i.e.}, an up and a down tetrahedron). All possible monopole arrangements in this minimum block and their degeneracies are depicted in Fig.~\ref{Fig1}b).
$ii-$ We define the field $\bm{\Sigma}$ equal to $\textbf{S}$ in alternate planes perpendicular to the [100] direction (black lines in Fig.~\ref{Fig1}), and to $-\textbf{S}$ in the others (red lines). We can check in Fig.~\ref{Fig1}b) that there is no net flux of $\bm{\Sigma}$ through the surface of any possible block. 
The same happens with the field $\bm{\Sigma}^{dim}$, represented by the dimer-spins (inverted or not depending on which [100] plane sits) and equal to zero in all the other sites.
It follows that any given configuration $\bm{\Sigma}$ and $\bm{\Sigma}^{dim}$ will have no net flux out of any closed surface including an integer number of blocks. The oblong Gaussian surface parallel to [100] drawn in green in Fig.~\ref{Fig1}c) is an example of this. If we neglect the outflow through the short walls, the condition for zero flux of $\bm{\Sigma}^{dim}$ is that the same number of dimer spins crosses the top and the bottom surface. Since each of them starts in a $-$ monopole and ends in a $+$ one, the \textit{total} charge in two alternate diamond [100] planes should be the same, assuring long range charge order along this direction. On top of this, the dimer spins propagate local monopole correlations, explaining the bow-tie shaped ``monopole" pinch points in Fig.~\ref{Fig2}a).
In a similar fashion, we can justify the ``spin" pinch points in Fig.~\ref{Fig2}b) integrating $\bm{\Sigma}$ along a vertical narrow Gaussian.

It is interesting to consider how magnetic charge fluctuations take place in the PML. Our conservation law implies that the total charge in a plane can only vary by the creation of \textit{stripes of monopoles} spanning the system along [100]. 
The monopole arrangement within a [100] plane has other type of fluctuations for a fixed total charge, since spin flipping along certain closed loops can preserve its overall charge. As in spin ices, and similar to the hexagons in ring exchange of quantum dimer models \cite{Hermele2004}, the spin loops turn into vortices if $\bm{\Sigma^{dim}}$ replaces $\bf{S}$.

\emph{Helmholtz decomposition in a monopole liquid.} 
%
% -----------  Iterative Decomposition ------------------------
Differently from the FCSL ---and in spite of their identical residual entropy--- spins in the PML phase are so correlated that for a given configuration of monopoles it is generally impossible for their magnetic moments to fluctuate.  Using our introductory analogy: if this were a fluid of magnetic ions, we can say that flipping the internal spin of an ion requires displacing a chain or loop of them: internal spin and charge are here aspects of the particle (in our case, a monopole) that cannot generally be separated. Due to this, we can now see that the unique monopole distribution behind Fig.~\ref{Fig2}a) is inherited from the spin correlations. The Helmholtz decomposition language can put light into this peculiarity. As mentioned in the introduction, the coherent part of the field \cite{Squires2012} in the case of the FCSL is given by $\textbf{S}_m$, and the diffuse one by $\textbf{S}_d$. Crucially, the diffuse field $\Delta\textbf{S}$ in the case of the PML comprehends \textit{both components}.

To our knowledge, the Helmholtz decomposition has only been tried on monopole matter for the FCSL~\cite{BBartlett2014,Jaubert2015}, where the (ordered) monopolar component $\textbf{S}_m$ can be determined from the start for all spin configurations. We have extended this to include monopole fluids, introducing an iterative method capable of obtaining the two components of any Ising configuration on a pyrochlore lattice to any degree of accuracy. This is an impossible task if we follow the recipe of Ref.~\citenum{BBartlett2014}~\footnote{This recipe would imply the need of an impossible field $\textbf{S}_m$, such that two tetrahedra sharing a vertex could have all their spins pointing in.}. We describe our method briefly in the following~\footnote{For more details on this method please see D. Slobinsky and R. A. Borzi, in preparation (2019).}.

We call $\textbf{S}_m^{(s)}$ and $\textbf{S}_d^{(s)}$ the modified field configurations at iteration step $s$, with $\textbf{S}_d^{(0)}=\textbf{S}$ and $\textbf{S}^{(0)}_m=0$. The method consists on sequentially subtracting a carefully chosen divergenceful 
field $\textbf{S}_{DF}^{(s+1)}$ (with a magnitude decreasing with $s$) from $\textbf{S}_d^{(s)}$, so as to render it divergenceless, and add it to $\textbf{S}_m^{(s)}$ ($\textbf{S}_m^{(s+1)} = \textbf{S}_m^{(s)} + \textbf{S}_{DF }^{(s+1)}$ and $\textbf{S}_d^{(s+1)} = \textbf{S}_d^{(s)} - \textbf{S}_{DF }^{(s+1)} $)
We define $\textbf{S}_{DF }^{(s+1)}$ as the superposition of configurations ``all-spins-in" or ``all-out"
taken independently in \textit{both up and down tetrahedra}; 
we chose the equal magnitude and sign of the four spins in a tetrahedron so that its associated monopole charge is half of that remaining in $\textbf{S}_d^{(s)}$ ($Q^{(s)}/2$). We found that the maximum monopole charge present in $\textbf{S}_d^{(s)}$ converges smoothly to zero with $s$, so that $\textbf{S}_d^{(s)} \rightarrow \textbf{S}_d$ when $s \to \infty$.

The neutron structure factors associated with the two field components so obtained (maximum charge left in the dipolar component $\textbf{S}_d^{(s)}$ less than $10^{-4}$, for $s\approx500$) is displayed in Fig. \ref{Fig2}c) and d). The Helmholtz decomposition makes now pristine that in the PML \textit{even the divergenceful component} corresponds to a Coulomb phase. 
Quite remarkably, the pinch points associated with $\textbf{S}_m^{(s)}$ are more intense in places where the AIAO double monopole crystal shows Bragg peaks, while those of $\textbf{S}_d^{(s)}$ concentrate in multiples of $\textbf{q}=$ [111] and [200].

%%%%%%%%%%%%%%%%%%%%%%%%%%%%%%%%%%%%%%%%%%%%%%%%%%%%%%%%%%%%%%%%%%%%%
%%%%%%%%%%%%%%%%%%%%%%%   STABILIZATION    %%%%%%%%%%%%%%%%%%%%%%%%%%
%%%%%%%%%%%%%%%%%%%%%%%%%%%%%%%%%%%%%%%%%%%%%%%%%%%%%%%%%%%%%%%%%%%%%
%%%%%%%%%%%%%%%%%%%%%%%%%%%%%%%%%%%%%%%%%%%%%%%%%%%%%%%%%%%%%%%%%%%%%
{\it Stabilization of the PML.}
While it is not clear how to stabilize the FCSL, 
the PML phase can be obtained in principle from an antiferromagnetic AIAO system by applying a magnetic field.
Different from Ref.~\cite{Guruciaga2016}, since $\textbf{H}~//$ [100] an uniaxial tensile stress in the same direction ($x$-axis) is needed to disfavor neutral tetrahedra configurations with net magnetic moment along $\textbf{H}$~\cite{Jaubert2010Multicriticality}. To show this, we use the nearest neighbors Hamiltonian

\begin{figure}
    \includegraphics[width=0.45\textwidth]{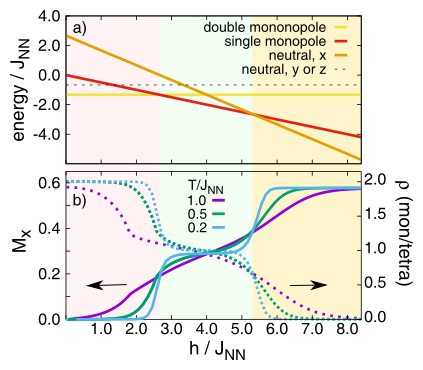}
    \caption{Realization of a \emph{Polarized Monopole Liquid} in an AIAO Ising pyrochlore under the action of a tetragonal distortion and a magnetic field along [100] ($x$-axis). 
    a) The energy per spin for a single tetrahedron vs. magnetic field ($J_{NN} = 0.06K$, $\delta = -0.1K$). On increasing fields we see three different ground states, respectively populated by 
    double monopoles, single monopoles with a positive component of magnetic moment along $x$ (the PML), and neutral tetrahedra with a positive component along $x$.  
    b) Monte Carlo simulations for Eq.~\ref{hamiltonian}, with the previous parameters and $\approx 8000$ spins. The magnetization $M_x$ and monopole density $\rho$ are fully consistent with a PML in the intermediate field interval.}
    \label{Fig3}
\end{figure}

\begin{equation}\label{hamiltonian}
     {\cal H}_{NN} =  \sum_{\langle i,j\rangle} (-J_{NN}+\epsilon_{ij}\delta) \sigma_i \sigma_j - h M_x, \, \, \, \, \, \, 
 \end{equation}

\noindent where $\epsilon_{ij}=0$ for links perpendicular to the \textit{x}-axis, and $-1$ for the others (see Fig.~\ref{Fig1} and Ref.~\citenum{Jaubert2010Multicriticality}); $h$ is the modulus of the field times the magnetic moment parallel to [100], $M_x$ the dimensionless magnetization along $x$, and $\sigma_i$ the spin ice pseudo-spins \cite{moessner1998rapcomm,Melko2004,Jaubert2008}. The new crucial parameter $\delta$ measures the relative change in the exchange constants due to uniaxial deformation. 
As exemplified in Fig. \ref{Fig3}a), with $\delta < 0$ and $0 < J_{NN} < |\delta|$, we can find a range of fields such that the stable ground state is populated only by the four types of single monopoles with positive magnetic moment components along [100], {\it i.e.}, a PML.
Our Monte Carlo simulations confirm this scenario (Fig.~\ref{Fig3}; see Sup. Mat.). We use $\delta = -0.1$ K, a value that can be imposed for example in an Ising pyrochlore like Dy$_2$Ti$_2$O$_7$~\cite{Mito2007} using piezoelectric devices \cite{Jaubert2010Multicriticality,Steppke2017}\footnote{Much bigger values can be reached depending on the progress of this technology~\cite{Hicks2014piezo} and the  magnetoelasticity of the material.}.
Fig. \ref{Fig3}b) shows that $M_x$ has three plateaus as a function of field, consistent with the three regions of stability seen in panel a). The $M_x$ value and the density of monopoles of the intermediate plateau reproduces those expected for a PML ($M_x/M_{sat} = 1/2$ and $\rho=1$). The calculated residual entropy for this phase (not shown) is compatible with $\frac{1}{4}\ln \left( \frac{27}{16} \right)$.
A conclusive experimental evidence~\footnote{Materials with $T_N$ as small as $\sim 0.3 K$ have been synthesized \cite{Petit2016,Lefrancois2017,Singh2008}. Their physics is more complicated than that described by the classical Hamiltonian in Eq.~\ref{hamiltonian}, but they are promising options to check under the proposed conditions.} of the PML phase could be reached studying the evolution of diffuse neutron scattering around $\textbf{q}=$[022]. 
For $|\delta|/J > 1$, diffuse scattering should develop in the form of \textit{pinch points} for the intermediate magnetic field window (see Fig. 3), before it disappears again at higher values.

% ------------- Dipolar perturbance ---------------
 The consideration of dipolar interactions poses extra challenges to the experimental observation of the PML, but also adds new possibilities. 
 %Balancing $J_{NN}$ and $\delta$, with
 For moderate $H$ and a small enough dipolar constant $D$~\cite{Melko2004}, we expect a cascade of phases (all within PML configurations) on decreasing temperature. The PML gives place to a crystal of single monopoles with fluctuating magnetic moments (a polarized version of the FCSL~\footnote{It is easy to check that, although degenerated, the entropy of this phase is subextensive}), and then these last degrees of freedom freeze. It is interesting to remember now that the monopolar degrees of freedom are described by a gauge field in the PML. Due to this, the melting of the monopole crystal into the PML can only proceed by introducing \textit{stripes of monopoles} spanning the system along [100] ---which are the excitations of the system in the charge channel--- in order to preserve the charge on alternate [100] planes. These non-local requirement on top of a solid-liquid transition may have an important effect on the nature of the phase transition~\cite{Moessner03,Nagle1989,Jaubert2008}.

%%%%%%%%%%%%%%%%%%%%%%%%%%%%%%%%%%%%%%%%%%%%%%%%%%%%%%%%%%%%%%%%%%%%%
%%%%%%%%%%%%%%%%%%%%%%     CONCLUSION     %%%%%%%%%%%%%%%%%%%%%%%%%%%
%%%%%%%%%%%%%%%%%%%%%%%%%%%%%%%%%%%%%%%%%%%%%%%%%%%%%%%%%%%%%%%%%%%%%
%%%%%%%%%%%%%%%%%%%%%%%%%%%%%%%%%%%%%%%%%%%%%%%%%%%%%%%%%%%%%%%%%%%%%
{\it Conclusions.}  
In summary, we have shown the existence of a new phase of monopole matter which is a liquid both in the structural (with monopoles as building blocks) and in the magnetic sense. 
Like in the \textit{fragmented Coulomb spin liquid}~\cite{BBartlett2014}, the Polarized Monopole Liquid hosts a Coulomb phase; however, here it is manifested in a very different way. Although both monopole states share the same residual entropy, the polarization of the PML couples the displacement of monopoles with their internal magnetic moments. These composite fluctuations are behind a PML's peculiarity: the algebraically decaying spin correlations propagate to the magnetic charge-charge correlations, leading to pinch points in the \textit{monopole} structure factor. These, in turn, imply that charge fluctuations within [100] planes could only occur through stripes of monopoles, spanning the entire system.
These peculiar features were further explored using a simple iterative method to obtain the Helmholtz decomposition of the lattice spin field. It showed that in the PML even the monopolar (divergenceful) component of the field has associated pinch points. Finally, we proposed how to identify in an experiment this peculiar phase, stabilized through the combined action of stress and magnetic field applied along [100].

%%%%%%%%%%%%%%%%%%%%%%%%%%%%%%%%%%%%%%%%%%%%%%%%%%%%%%%%%%%%%%%%%%%%%
%%%%%%%%%%%%%%%%%%%%%%% Acknowledgments %%%%%%%%%%%%%%%%%%%%%%%%%%%%%
%%%%%%%%%%%%%%%%%%%%%%%%%%%%%%%%%%%%%%%%%%%%%%%%%%%%%%%%%%%%%%%%%%%%%
%%%%%%%%%%%%%%%%%%%%%%%%%%%%%%%%%%%%%%%%%%%%%%%%%%%%%%%%%%%%%%%%%%%%%
\begin{acknowledgments}
The authors acknowledge financial support from ANPCyT (PICT 2013 N$^{\circ}$2004, PICT 2014 N$^{\circ}$2618), and CONICET (PIP 0446).
\end{acknowledgments}

\bibliography{references}

\begin{thebibliography}{53}%
\makeatletter
\providecommand \@ifxundefined [1]{%
 \@ifx{#1\undefined}
}%
\providecommand \@ifnum [1]{%
 \ifnum #1\expandafter \@firstoftwo
 \else \expandafter \@secondoftwo
 \fi
}%
\providecommand \@ifx [1]{%
 \ifx #1\expandafter \@firstoftwo
 \else \expandafter \@secondoftwo
 \fi
}%
\providecommand \natexlab [1]{#1}%
\providecommand \enquote  [1]{``#1''}%
\providecommand \bibnamefont  [1]{#1}%
\providecommand \bibfnamefont [1]{#1}%
\providecommand \citenamefont [1]{#1}%
\providecommand \href@noop [0]{\@secondoftwo}%
\providecommand \href [0]{\begingroup \@sanitize@url \@href}%
\providecommand \@href[1]{\@@startlink{#1}\@@href}%
\providecommand \@@href[1]{\endgroup#1\@@endlink}%
\providecommand \@sanitize@url [0]{\catcode `\\12\catcode `\$12\catcode
  `\&12\catcode `\#12\catcode `\^12\catcode `\_12\catcode `\%12\relax}%
\providecommand \@@startlink[1]{}%
\providecommand \@@endlink[0]{}%
\providecommand \url  [0]{\begingroup\@sanitize@url \@url }%
\providecommand \@url [1]{\endgroup\@href {#1}{\urlprefix }}%
\providecommand \urlprefix  [0]{URL }%
\providecommand \Eprint [0]{\href }%
\providecommand \doibase [0]{http://dx.doi.org/}%
\providecommand \selectlanguage [0]{\@gobble}%
\providecommand \bibinfo  [0]{\@secondoftwo}%
\providecommand \bibfield  [0]{\@secondoftwo}%
\providecommand \translation [1]{[#1]}%
\providecommand \BibitemOpen [0]{}%
\providecommand \bibitemStop [0]{}%
\providecommand \bibitemNoStop [0]{.\EOS\space}%
\providecommand \EOS [0]{\spacefactor3000\relax}%
\providecommand \BibitemShut  [1]{\csname bibitem#1\endcsname}%
\let\auto@bib@innerbib\@empty
%</preamble>
\bibitem [{\citenamefont {Ramirez}(1994)}]{Ramirez1994}%
  \BibitemOpen
  \bibfield  {author} {\bibinfo {author} {\bibfnamefont {A.}~\bibnamefont
  {Ramirez}},\ }\href@noop {} {\bibfield  {journal} {\bibinfo  {journal}
  {Annual Review of Materials Science}\ }\textbf {\bibinfo {volume} {24}},\
  \bibinfo {pages} {453} (\bibinfo {year} {1994})}\BibitemShut {NoStop}%
\bibitem [{\citenamefont {Ramirez}\ \emph {et~al.}(1999)\citenamefont
  {Ramirez}, \citenamefont {Hayashi}, \citenamefont {Cava}, \citenamefont
  {Siddharthan},\ and\ \citenamefont {Shastry}}]{Ramirez1999}%
  \BibitemOpen
  \bibfield  {author} {\bibinfo {author} {\bibfnamefont {A.~P.}\ \bibnamefont
  {Ramirez}}, \bibinfo {author} {\bibfnamefont {A.}~\bibnamefont {Hayashi}},
  \bibinfo {author} {\bibfnamefont {R.}~\bibnamefont {Cava}}, \bibinfo {author}
  {\bibfnamefont {R.}~\bibnamefont {Siddharthan}}, \ and\ \bibinfo {author}
  {\bibfnamefont {B.}~\bibnamefont {Shastry}},\ }\href@noop {} {\bibfield
  {journal} {\bibinfo  {journal} {Nature}\ }\textbf {\bibinfo {volume} {399}},\
  \bibinfo {pages} {333} (\bibinfo {year} {1999})}\BibitemShut {NoStop}%
\bibitem [{\citenamefont {Castelnovo}\ \emph {et~al.}(2008)\citenamefont
  {Castelnovo}, \citenamefont {Moessner},\ and\ \citenamefont
  {Sondhi}}]{Castelnovo2008}%
  \BibitemOpen
  \bibfield  {author} {\bibinfo {author} {\bibfnamefont {C.}~\bibnamefont
  {Castelnovo}}, \bibinfo {author} {\bibfnamefont {R.}~\bibnamefont
  {Moessner}}, \ and\ \bibinfo {author} {\bibfnamefont {S.~L.}\ \bibnamefont
  {Sondhi}},\ }\href {\doibase 10.1038/nature06433} {\bibfield  {journal}
  {\bibinfo  {journal} {Nature Physics}\ }\textbf {\bibinfo {volume} {451}},\
  \bibinfo {pages} {42} (\bibinfo {year} {2008})}\BibitemShut {NoStop}%
\bibitem [{\citenamefont {Morris}\ \emph {et~al.}(2009)\citenamefont {Morris},
  \citenamefont {Tennant}, \citenamefont {Grigera}, \citenamefont {Klemke},
  \citenamefont {Castelnovo}, \citenamefont {Moessner}, \citenamefont
  {Czternasty}, \citenamefont {Meissner}, \citenamefont {Rule}, \citenamefont
  {Hoffmann}, \citenamefont {Kiefer}, \citenamefont {Gerischer}, \citenamefont
  {Slobinsky},\ and\ \citenamefont {Perry}}]{Morris2009}%
  \BibitemOpen
  \bibfield  {author} {\bibinfo {author} {\bibfnamefont {D.~J.~P.}\
  \bibnamefont {Morris}}, \bibinfo {author} {\bibfnamefont {D.~A.}\
  \bibnamefont {Tennant}}, \bibinfo {author} {\bibfnamefont {S.~A.}\
  \bibnamefont {Grigera}}, \bibinfo {author} {\bibfnamefont {B.}~\bibnamefont
  {Klemke}}, \bibinfo {author} {\bibfnamefont {C.}~\bibnamefont {Castelnovo}},
  \bibinfo {author} {\bibfnamefont {R.}~\bibnamefont {Moessner}}, \bibinfo
  {author} {\bibfnamefont {C.}~\bibnamefont {Czternasty}}, \bibinfo {author}
  {\bibfnamefont {M.}~\bibnamefont {Meissner}}, \bibinfo {author}
  {\bibfnamefont {K.~C.}\ \bibnamefont {Rule}}, \bibinfo {author}
  {\bibfnamefont {J.-U.}\ \bibnamefont {Hoffmann}}, \bibinfo {author}
  {\bibfnamefont {K.}~\bibnamefont {Kiefer}}, \bibinfo {author} {\bibfnamefont
  {S.}~\bibnamefont {Gerischer}}, \bibinfo {author} {\bibfnamefont
  {D.}~\bibnamefont {Slobinsky}}, \ and\ \bibinfo {author} {\bibfnamefont
  {R.~S.}\ \bibnamefont {Perry}},\ }\href {\doibase 10.1126/science.1178868}
  {\bibfield  {journal} {\bibinfo  {journal} {Science}\ }\textbf {\bibinfo
  {volume} {326}},\ \bibinfo {pages} {411} (\bibinfo {year}
  {2009})}\BibitemShut {NoStop}%
\bibitem [{\citenamefont {Jaubert}\ and\ \citenamefont
  {Holdsworth}(2009)}]{jaubert2009nat}%
  \BibitemOpen
  \bibfield  {author} {\bibinfo {author} {\bibfnamefont {L.~D.}\ \bibnamefont
  {Jaubert}}\ and\ \bibinfo {author} {\bibfnamefont {P.~C.}\ \bibnamefont
  {Holdsworth}},\ }\href@noop {} {\bibfield  {journal} {\bibinfo  {journal}
  {Nature Physics}\ }\textbf {\bibinfo {volume} {5}},\ \bibinfo {pages} {258}
  (\bibinfo {year} {2009})}\BibitemShut {NoStop}%
\bibitem [{\citenamefont {Chern}\ \emph {et~al.}(2011)\citenamefont {Chern},
  \citenamefont {Mellado},\ and\ \citenamefont {Tchernyshyov}}]{Tch2011}%
  \BibitemOpen
  \bibfield  {author} {\bibinfo {author} {\bibfnamefont {G.-W.}\ \bibnamefont
  {Chern}}, \bibinfo {author} {\bibfnamefont {P.}~\bibnamefont {Mellado}}, \
  and\ \bibinfo {author} {\bibfnamefont {O.}~\bibnamefont {Tchernyshyov}},\
  }\href {\doibase 10.1103/PhysRevLett.106.207202} {\bibfield  {journal}
  {\bibinfo  {journal} {Phys. Rev. Lett.}\ }\textbf {\bibinfo {volume} {106}},\
  \bibinfo {pages} {207202} (\bibinfo {year} {2011})}\BibitemShut {NoStop}%
\bibitem [{\citenamefont {Sazonov}\ \emph {et~al.}(2012)\citenamefont
  {Sazonov}, \citenamefont {Gukasov}, \citenamefont {Mirebeau},\ and\
  \citenamefont {Bonville}}]{Sazonov2012}%
  \BibitemOpen
  \bibfield  {author} {\bibinfo {author} {\bibfnamefont {A.~P.}\ \bibnamefont
  {Sazonov}}, \bibinfo {author} {\bibfnamefont {A.}~\bibnamefont {Gukasov}},
  \bibinfo {author} {\bibfnamefont {I.}~\bibnamefont {Mirebeau}}, \ and\
  \bibinfo {author} {\bibfnamefont {P.}~\bibnamefont {Bonville}},\ }\href
  {\doibase 10.1103/PhysRevB.85.214420} {\bibfield  {journal} {\bibinfo
  {journal} {Phys. Rev. B}\ }\textbf {\bibinfo {volume} {85}},\ \bibinfo
  {pages} {214420} (\bibinfo {year} {2012})}\BibitemShut {NoStop}%
\bibitem [{\citenamefont {Borzi}\ \emph {et~al.}(2013)\citenamefont {Borzi},
  \citenamefont {Slobinsky},\ and\ \citenamefont {Grigera}}]{Borzi2013}%
  \BibitemOpen
  \bibfield  {author} {\bibinfo {author} {\bibfnamefont {R.~A.}\ \bibnamefont
  {Borzi}}, \bibinfo {author} {\bibfnamefont {D.}~\bibnamefont {Slobinsky}}, \
  and\ \bibinfo {author} {\bibfnamefont {S.~A.}\ \bibnamefont {Grigera}},\
  }\href {\doibase 10.1103/PhysRevLett.111.147204} {\bibfield  {journal}
  {\bibinfo  {journal} {Phys. Rev. Lett.}\ }\textbf {\bibinfo {volume} {111}},\
  \bibinfo {pages} {147204} (\bibinfo {year} {2013})}\BibitemShut {NoStop}%
\bibitem [{\citenamefont {Brooks-Bartlett}\ \emph {et~al.}(2014)\citenamefont
  {Brooks-Bartlett}, \citenamefont {Banks}, \citenamefont {Jaubert},
  \citenamefont {Harman-Clarke},\ and\ \citenamefont
  {Holdsworth}}]{BBartlett2014}%
  \BibitemOpen
  \bibfield  {author} {\bibinfo {author} {\bibfnamefont {M.~E.}\ \bibnamefont
  {Brooks-Bartlett}}, \bibinfo {author} {\bibfnamefont {S.~T.}\ \bibnamefont
  {Banks}}, \bibinfo {author} {\bibfnamefont {L.~D.~C.}\ \bibnamefont
  {Jaubert}}, \bibinfo {author} {\bibfnamefont {A.}~\bibnamefont
  {Harman-Clarke}}, \ and\ \bibinfo {author} {\bibfnamefont {P.~C.~W.}\
  \bibnamefont {Holdsworth}},\ }\href {\doibase 10.1103/PhysRevX.4.011007}
  {\bibfield  {journal} {\bibinfo  {journal} {Phys. Rev. X}\ }\textbf {\bibinfo
  {volume} {4}},\ \bibinfo {pages} {011007} (\bibinfo {year}
  {2014})}\BibitemShut {NoStop}%
\bibitem [{\citenamefont {Guruciaga}\ \emph {et~al.}(2014)\citenamefont
  {Guruciaga}, \citenamefont {Grigera},\ and\ \citenamefont
  {Borzi}}]{Guruciaga2014}%
  \BibitemOpen
  \bibfield  {author} {\bibinfo {author} {\bibfnamefont {P.~C.}\ \bibnamefont
  {Guruciaga}}, \bibinfo {author} {\bibfnamefont {S.~A.}\ \bibnamefont
  {Grigera}}, \ and\ \bibinfo {author} {\bibfnamefont {R.~A.}\ \bibnamefont
  {Borzi}},\ }\href@noop {} {\bibfield  {journal} {\bibinfo  {journal}
  {Physical Review B}\ }\textbf {\bibinfo {volume} {90}},\ \bibinfo {pages}
  {184423} (\bibinfo {year} {2014})}\BibitemShut {NoStop}%
\bibitem [{\citenamefont {Jaubert}\ and\ \citenamefont
  {Moessner}(2015)}]{jaubert2015p}%
  \BibitemOpen
  \bibfield  {author} {\bibinfo {author} {\bibfnamefont {L.~D.~C.}\
  \bibnamefont {Jaubert}}\ and\ \bibinfo {author} {\bibfnamefont
  {R.}~\bibnamefont {Moessner}},\ }\href {\doibase 10.1103/PhysRevB.91.214422}
  {\bibfield  {journal} {\bibinfo  {journal} {Phys. Rev. B}\ }\textbf {\bibinfo
  {volume} {91}},\ \bibinfo {pages} {214422} (\bibinfo {year}
  {2015})}\BibitemShut {NoStop}%
\bibitem [{\citenamefont {{Rau Jeffrey G.}}\ and\ \citenamefont {{Gingras
  Michel J. P.}}(2016)}]{Rau2016}%
  \BibitemOpen
  \bibfield  {author} {\bibinfo {author} {\bibnamefont {{Rau Jeffrey G.}}}\
  and\ \bibinfo {author} {\bibnamefont {{Gingras Michel J. P.}}},\ }\href
  {\doibase http://dx.doi.org/10.1038/ncomms12234 10.1038/ncomms12234}
  {\bibfield  {journal} {\bibinfo  {journal} {Nature Communications}\ }\textbf
  {\bibinfo {volume} {7}},\ \bibinfo {pages} {12234} (\bibinfo {year}
  {2016})}\BibitemShut {NoStop}%
\bibitem [{\citenamefont {Udagawa}\ \emph {et~al.}(2016)\citenamefont
  {Udagawa}, \citenamefont {Jaubert}, \citenamefont {Castelnovo},\ and\
  \citenamefont {Moessner}}]{Udagawa2016}%
  \BibitemOpen
  \bibfield  {author} {\bibinfo {author} {\bibfnamefont {M.}~\bibnamefont
  {Udagawa}}, \bibinfo {author} {\bibfnamefont {L.}~\bibnamefont {Jaubert}},
  \bibinfo {author} {\bibfnamefont {C.}~\bibnamefont {Castelnovo}}, \ and\
  \bibinfo {author} {\bibfnamefont {R.}~\bibnamefont {Moessner}},\ }\href@noop
  {} {\bibfield  {journal} {\bibinfo  {journal} {Physical Review B}\ }\textbf
  {\bibinfo {volume} {94}},\ \bibinfo {pages} {104416} (\bibinfo {year}
  {2016})}\BibitemShut {NoStop}%
\bibitem [{\citenamefont {Jaubert}(2015)}]{Jaubert2015}%
  \BibitemOpen
  \bibfield  {author} {\bibinfo {author} {\bibfnamefont {L.~D.~C.}\
  \bibnamefont {Jaubert}},\ }\href {\doibase 10.1142/S2010324715400056}
  {\bibfield  {journal} {\bibinfo  {journal} {SPIN}\ }\textbf {\bibinfo
  {volume} {05}},\ \bibinfo {pages} {1540005} (\bibinfo {year}
  {2015})}\BibitemShut {NoStop}%
\bibitem [{Note1()}]{Note1}%
  \BibitemOpen
  \bibinfo {note} {In the general case, in addition to the previous fields, a
  harmonic contribution $\protect \textbf {S}_{har}$ can also be present \cite
  {Bramwell2017}.}\BibitemShut {Stop}%
\bibitem [{\citenamefont {Henley}(2010)}]{Henley2010}%
  \BibitemOpen
  \bibfield  {author} {\bibinfo {author} {\bibfnamefont {C.~L.}\ \bibnamefont
  {Henley}},\ }\href {\doibase 10.1146/annurev-conmatphys-070909-104138}
  {\bibfield  {journal} {\bibinfo  {journal} {Annual Review of Condensed Matter
  Physics}\ }\textbf {\bibinfo {volume} {1}},\ \bibinfo {pages} {179} (\bibinfo
  {year} {2010})}\BibitemShut {NoStop}%
\bibitem [{\citenamefont {Moessner}\ and\ \citenamefont
  {Chalker}(1998)}]{Moessner1998}%
  \BibitemOpen
  \bibfield  {author} {\bibinfo {author} {\bibfnamefont {R.}~\bibnamefont
  {Moessner}}\ and\ \bibinfo {author} {\bibfnamefont {J.~T.}\ \bibnamefont
  {Chalker}},\ }\href {\doibase 10.1103/PhysRevB.58.12049} {\bibfield
  {journal} {\bibinfo  {journal} {Phys. Rev. B}\ }\textbf {\bibinfo {volume}
  {58}},\ \bibinfo {pages} {12049} (\bibinfo {year} {1998})}\BibitemShut
  {NoStop}%
\bibitem [{\citenamefont {Fennell}\ \emph {et~al.}(2007)\citenamefont
  {Fennell}, \citenamefont {Bramwell}, \citenamefont {McMorrow}, \citenamefont
  {Manuel},\ and\ \citenamefont {Wildes}}]{Fennell2007}%
  \BibitemOpen
  \bibfield  {author} {\bibinfo {author} {\bibfnamefont {T.}~\bibnamefont
  {Fennell}}, \bibinfo {author} {\bibfnamefont {S.}~\bibnamefont {Bramwell}},
  \bibinfo {author} {\bibfnamefont {D.}~\bibnamefont {McMorrow}}, \bibinfo
  {author} {\bibfnamefont {P.}~\bibnamefont {Manuel}}, \ and\ \bibinfo {author}
  {\bibfnamefont {A.}~\bibnamefont {Wildes}},\ }\href@noop {} {\bibfield
  {journal} {\bibinfo  {journal} {Nature Physics}\ }\textbf {\bibinfo {volume}
  {3}},\ \bibinfo {pages} {566} (\bibinfo {year} {2007})}\BibitemShut {NoStop}%
\bibitem [{\citenamefont {Fennell}\ \emph {et~al.}(2009)\citenamefont
  {Fennell}, \citenamefont {Deen}, \citenamefont {Wildes}, \citenamefont
  {Schmalzl}, \citenamefont {Prabhakaran}, \citenamefont {Boothroyd},
  \citenamefont {Aldus}, \citenamefont {McMorrow},\ and\ \citenamefont
  {Bramwell}}]{Fennell2009}%
  \BibitemOpen
  \bibfield  {author} {\bibinfo {author} {\bibfnamefont {T.}~\bibnamefont
  {Fennell}}, \bibinfo {author} {\bibfnamefont {P.~P.}\ \bibnamefont {Deen}},
  \bibinfo {author} {\bibfnamefont {A.~R.}\ \bibnamefont {Wildes}}, \bibinfo
  {author} {\bibfnamefont {K.}~\bibnamefont {Schmalzl}}, \bibinfo {author}
  {\bibfnamefont {D.}~\bibnamefont {Prabhakaran}}, \bibinfo {author}
  {\bibfnamefont {A.~T.}\ \bibnamefont {Boothroyd}}, \bibinfo {author}
  {\bibfnamefont {R.~J.}\ \bibnamefont {Aldus}}, \bibinfo {author}
  {\bibfnamefont {D.~F.}\ \bibnamefont {McMorrow}}, \ and\ \bibinfo {author}
  {\bibfnamefont {S.~T.}\ \bibnamefont {Bramwell}},\ }\href {\doibase
  10.1126/science.1177582} {\bibfield  {journal} {\bibinfo  {journal}
  {Science}\ }\textbf {\bibinfo {volume} {326}},\ \bibinfo {pages} {415}
  (\bibinfo {year} {2009})}\BibitemShut {NoStop}%
\bibitem [{\citenamefont {Chang}\ \emph {et~al.}(2010)\citenamefont {Chang},
  \citenamefont {Su}, \citenamefont {Kao}, \citenamefont {Chou}, \citenamefont
  {Mittal}, \citenamefont {Schneider}, \citenamefont {Br\"uckel}, \citenamefont
  {Balakrishnan},\ and\ \citenamefont {Lees}}]{Chang2010}%
  \BibitemOpen
  \bibfield  {author} {\bibinfo {author} {\bibfnamefont {L.~J.}\ \bibnamefont
  {Chang}}, \bibinfo {author} {\bibfnamefont {Y.}~\bibnamefont {Su}}, \bibinfo
  {author} {\bibfnamefont {Y.-J.}\ \bibnamefont {Kao}}, \bibinfo {author}
  {\bibfnamefont {Y.~Z.}\ \bibnamefont {Chou}}, \bibinfo {author}
  {\bibfnamefont {R.}~\bibnamefont {Mittal}}, \bibinfo {author} {\bibfnamefont
  {H.}~\bibnamefont {Schneider}}, \bibinfo {author} {\bibfnamefont
  {T.}~\bibnamefont {Br\"uckel}}, \bibinfo {author} {\bibfnamefont
  {G.}~\bibnamefont {Balakrishnan}}, \ and\ \bibinfo {author} {\bibfnamefont
  {M.~R.}\ \bibnamefont {Lees}},\ }\href {\doibase 10.1103/PhysRevB.82.172403}
  {\bibfield  {journal} {\bibinfo  {journal} {Phys. Rev. B}\ }\textbf {\bibinfo
  {volume} {82}},\ \bibinfo {pages} {172403} (\bibinfo {year}
  {2010})}\BibitemShut {NoStop}%
\bibitem [{\citenamefont {Fennell}\ \emph {et~al.}(2012)\citenamefont
  {Fennell}, \citenamefont {Kenzelmann}, \citenamefont {Roessli}, \citenamefont
  {Haas},\ and\ \citenamefont {Cava}}]{Fennell2012}%
  \BibitemOpen
  \bibfield  {author} {\bibinfo {author} {\bibfnamefont {T.}~\bibnamefont
  {Fennell}}, \bibinfo {author} {\bibfnamefont {M.}~\bibnamefont {Kenzelmann}},
  \bibinfo {author} {\bibfnamefont {B.}~\bibnamefont {Roessli}}, \bibinfo
  {author} {\bibfnamefont {M.~K.}\ \bibnamefont {Haas}}, \ and\ \bibinfo
  {author} {\bibfnamefont {R.~J.}\ \bibnamefont {Cava}},\ }\href {\doibase
  10.1103/PhysRevLett.109.017201} {\bibfield  {journal} {\bibinfo  {journal}
  {Phys. Rev. Lett.}\ }\textbf {\bibinfo {volume} {109}},\ \bibinfo {pages}
  {017201} (\bibinfo {year} {2012})}\BibitemShut {NoStop}%
\bibitem [{\citenamefont {Petit}\ \emph {et~al.}(2012)\citenamefont {Petit},
  \citenamefont {Bonville}, \citenamefont {Robert}, \citenamefont {Decorse},\
  and\ \citenamefont {Mirebeau}}]{Petit2012}%
  \BibitemOpen
  \bibfield  {author} {\bibinfo {author} {\bibfnamefont {S.}~\bibnamefont
  {Petit}}, \bibinfo {author} {\bibfnamefont {P.}~\bibnamefont {Bonville}},
  \bibinfo {author} {\bibfnamefont {J.}~\bibnamefont {Robert}}, \bibinfo
  {author} {\bibfnamefont {C.}~\bibnamefont {Decorse}}, \ and\ \bibinfo
  {author} {\bibfnamefont {I.}~\bibnamefont {Mirebeau}},\ }\href {\doibase
  10.1103/PhysRevB.86.174403} {\bibfield  {journal} {\bibinfo  {journal} {Phys.
  Rev. B}\ }\textbf {\bibinfo {volume} {86}},\ \bibinfo {pages} {174403}
  (\bibinfo {year} {2012})}\BibitemShut {NoStop}%
\bibitem [{\citenamefont {Petit}\ \emph {et~al.}(2016)\citenamefont {Petit},
  \citenamefont {Lhotel}, \citenamefont {Canals}, \citenamefont
  {Ciomaga-Hatnean}, \citenamefont {Ollivier}, \citenamefont {Mutka},
  \citenamefont {Ressouche}, \citenamefont {Wildes}, \citenamefont {Lees},\
  and\ \citenamefont {Balakrishnan}}]{Petit2016}%
  \BibitemOpen
  \bibfield  {author} {\bibinfo {author} {\bibfnamefont {S.}~\bibnamefont
  {Petit}}, \bibinfo {author} {\bibfnamefont {E.}~\bibnamefont {Lhotel}},
  \bibinfo {author} {\bibfnamefont {B.}~\bibnamefont {Canals}}, \bibinfo
  {author} {\bibfnamefont {M.}~\bibnamefont {Ciomaga-Hatnean}}, \bibinfo
  {author} {\bibfnamefont {J.}~\bibnamefont {Ollivier}}, \bibinfo {author}
  {\bibfnamefont {H.}~\bibnamefont {Mutka}}, \bibinfo {author} {\bibfnamefont
  {E.}~\bibnamefont {Ressouche}}, \bibinfo {author} {\bibfnamefont {A.~R.}\
  \bibnamefont {Wildes}}, \bibinfo {author} {\bibfnamefont {M.~R.}\
  \bibnamefont {Lees}}, \ and\ \bibinfo {author} {\bibfnamefont
  {G.}~\bibnamefont {Balakrishnan}},\ }\href@noop {} {\bibfield  {journal}
  {\bibinfo  {journal} {Nature Physics}\ }\textbf {\bibinfo {volume} {12}},\
  \bibinfo {pages} {746} (\bibinfo {year} {2016})}\BibitemShut {NoStop}%
\bibitem [{\citenamefont {Lefran{\c{c}}ois}\ \emph {et~al.}(2017)\citenamefont
  {Lefran{\c{c}}ois}, \citenamefont {Cathelin}, \citenamefont {Lhotel},
  \citenamefont {Robert}, \citenamefont {Lejay}, \citenamefont {Colin},
  \citenamefont {Canals}, \citenamefont {Damay}, \citenamefont {Ollivier},
  \citenamefont {F{\aa}k} \emph {et~al.}}]{Lefrancois2017}%
  \BibitemOpen
  \bibfield  {author} {\bibinfo {author} {\bibfnamefont {E.}~\bibnamefont
  {Lefran{\c{c}}ois}}, \bibinfo {author} {\bibfnamefont {V.}~\bibnamefont
  {Cathelin}}, \bibinfo {author} {\bibfnamefont {E.}~\bibnamefont {Lhotel}},
  \bibinfo {author} {\bibfnamefont {J.}~\bibnamefont {Robert}}, \bibinfo
  {author} {\bibfnamefont {P.}~\bibnamefont {Lejay}}, \bibinfo {author}
  {\bibfnamefont {C.~V.}\ \bibnamefont {Colin}}, \bibinfo {author}
  {\bibfnamefont {B.}~\bibnamefont {Canals}}, \bibinfo {author} {\bibfnamefont
  {F.}~\bibnamefont {Damay}}, \bibinfo {author} {\bibfnamefont
  {J.}~\bibnamefont {Ollivier}}, \bibinfo {author} {\bibfnamefont
  {B.}~\bibnamefont {F{\aa}k}},  \emph {et~al.},\ }\href@noop {} {\bibfield
  {journal} {\bibinfo  {journal} {Nature communications}\ }\textbf {\bibinfo
  {volume} {8}},\ \bibinfo {pages} {209} (\bibinfo {year} {2017})}\BibitemShut
  {NoStop}%
\bibitem [{\citenamefont {Canals}\ \emph {et~al.}(2016)\citenamefont {Canals},
  \citenamefont {Chioar}, \citenamefont {Nguyen}, \citenamefont {Hehn},
  \citenamefont {Lacour}, \citenamefont {Montaigne}, \citenamefont {Locatelli},
  \citenamefont {Mente{\c{s}}}, \citenamefont {Burgos},\ and\ \citenamefont
  {Rougemaille}}]{Canals2016}%
  \BibitemOpen
  \bibfield  {author} {\bibinfo {author} {\bibfnamefont {B.}~\bibnamefont
  {Canals}}, \bibinfo {author} {\bibfnamefont {I.-A.}\ \bibnamefont {Chioar}},
  \bibinfo {author} {\bibfnamefont {V.-D.}\ \bibnamefont {Nguyen}}, \bibinfo
  {author} {\bibfnamefont {M.}~\bibnamefont {Hehn}}, \bibinfo {author}
  {\bibfnamefont {D.}~\bibnamefont {Lacour}}, \bibinfo {author} {\bibfnamefont
  {F.}~\bibnamefont {Montaigne}}, \bibinfo {author} {\bibfnamefont
  {A.}~\bibnamefont {Locatelli}}, \bibinfo {author} {\bibfnamefont {T.~O.}\
  \bibnamefont {Mente{\c{s}}}}, \bibinfo {author} {\bibfnamefont {B.~S.}\
  \bibnamefont {Burgos}}, \ and\ \bibinfo {author} {\bibfnamefont
  {N.}~\bibnamefont {Rougemaille}},\ }\href@noop {} {\bibfield  {journal}
  {\bibinfo  {journal} {Nature communications}\ }\textbf {\bibinfo {volume}
  {7}},\ \bibinfo {pages} {11446} (\bibinfo {year} {2016})}\BibitemShut
  {NoStop}%
\bibitem [{\citenamefont {Slobinsky}\ \emph {et~al.}(2018)\citenamefont
  {Slobinsky}, \citenamefont {Baglietto},\ and\ \citenamefont
  {Borzi}}]{Slo2018}%
  \BibitemOpen
  \bibfield  {author} {\bibinfo {author} {\bibfnamefont {D.}~\bibnamefont
  {Slobinsky}}, \bibinfo {author} {\bibfnamefont {G.}~\bibnamefont
  {Baglietto}}, \ and\ \bibinfo {author} {\bibfnamefont {R.~A.}\ \bibnamefont
  {Borzi}},\ }\href {\doibase 10.1103/PhysRevB.97.174422} {\bibfield  {journal}
  {\bibinfo  {journal} {Phys. Rev. B}\ }\textbf {\bibinfo {volume} {97}},\
  \bibinfo {pages} {174422} (\bibinfo {year} {2018})}\BibitemShut {NoStop}%
\bibitem [{\citenamefont {Nagle}(1966)}]{Nagle1966}%
  \BibitemOpen
  \bibfield  {author} {\bibinfo {author} {\bibfnamefont {J.~F.}\ \bibnamefont
  {Nagle}},\ }\href {\doibase 10.1103/PhysRev.152.190} {\bibfield  {journal}
  {\bibinfo  {journal} {Phys. Rev.}\ }\textbf {\bibinfo {volume} {152}},\
  \bibinfo {pages} {190} (\bibinfo {year} {1966})}\BibitemShut {NoStop}%
\bibitem [{Note2()}]{Note2}%
  \BibitemOpen
  \bibinfo {note} {This case of a liquid and a crystal having similar entropies
  resembles that of $^3$He under pressure.}\BibitemShut {Stop}%
\bibitem [{\citenamefont {Ueda}\ \emph {et~al.}(2005)\citenamefont {Ueda},
  \citenamefont {Katori}, \citenamefont {Mitamura}, \citenamefont {Goto},\ and\
  \citenamefont {Takagi}}]{Ueda2005}%
  \BibitemOpen
  \bibfield  {author} {\bibinfo {author} {\bibfnamefont {H.}~\bibnamefont
  {Ueda}}, \bibinfo {author} {\bibfnamefont {H.~A.}\ \bibnamefont {Katori}},
  \bibinfo {author} {\bibfnamefont {H.}~\bibnamefont {Mitamura}}, \bibinfo
  {author} {\bibfnamefont {T.}~\bibnamefont {Goto}}, \ and\ \bibinfo {author}
  {\bibfnamefont {H.}~\bibnamefont {Takagi}},\ }\href@noop {} {\bibfield
  {journal} {\bibinfo  {journal} {Physical review letters}\ }\textbf {\bibinfo
  {volume} {94}},\ \bibinfo {pages} {047202} (\bibinfo {year}
  {2005})}\BibitemShut {NoStop}%
\bibitem [{\citenamefont {Bergman}\ \emph {et~al.}(2006)\citenamefont
  {Bergman}, \citenamefont {Shindou}, \citenamefont {Fiete},\ and\
  \citenamefont {Balents}}]{Bergman2006b}%
  \BibitemOpen
  \bibfield  {author} {\bibinfo {author} {\bibfnamefont {D.~L.}\ \bibnamefont
  {Bergman}}, \bibinfo {author} {\bibfnamefont {R.}~\bibnamefont {Shindou}},
  \bibinfo {author} {\bibfnamefont {G.~A.}\ \bibnamefont {Fiete}}, \ and\
  \bibinfo {author} {\bibfnamefont {L.}~\bibnamefont {Balents}},\ }\href@noop
  {} {\bibfield  {journal} {\bibinfo  {journal} {Physical review letters}\
  }\textbf {\bibinfo {volume} {96}},\ \bibinfo {pages} {097207} (\bibinfo
  {year} {2006})}\BibitemShut {NoStop}%
\bibitem [{\citenamefont {Sikora}\ \emph {et~al.}(2011)\citenamefont {Sikora},
  \citenamefont {Shannon}, \citenamefont {Pollmann}, \citenamefont {Penc},\
  and\ \citenamefont {Fulde}}]{Sikora2011}%
  \BibitemOpen
  \bibfield  {author} {\bibinfo {author} {\bibfnamefont {O.}~\bibnamefont
  {Sikora}}, \bibinfo {author} {\bibfnamefont {N.}~\bibnamefont {Shannon}},
  \bibinfo {author} {\bibfnamefont {F.}~\bibnamefont {Pollmann}}, \bibinfo
  {author} {\bibfnamefont {K.}~\bibnamefont {Penc}}, \ and\ \bibinfo {author}
  {\bibfnamefont {P.}~\bibnamefont {Fulde}},\ }\href@noop {} {\bibfield
  {journal} {\bibinfo  {journal} {Physical Review B}\ }\textbf {\bibinfo
  {volume} {84}},\ \bibinfo {pages} {115129} (\bibinfo {year}
  {2011})}\BibitemShut {NoStop}%
\bibitem [{Note3()}]{Note3}%
  \BibitemOpen
  \bibinfo {note} {See Supplemental Material.}\BibitemShut {Stop}%
\bibitem [{\citenamefont {Bramwell}\ and\ \citenamefont
  {Gingras}(2001)}]{Bramwell2001}%
  \BibitemOpen
  \bibfield  {author} {\bibinfo {author} {\bibfnamefont {S.~T.}\ \bibnamefont
  {Bramwell}}\ and\ \bibinfo {author} {\bibfnamefont {M.~J.}\ \bibnamefont
  {Gingras}},\ }\href@noop {} {\bibfield  {journal} {\bibinfo  {journal}
  {Science}\ }\textbf {\bibinfo {volume} {294}},\ \bibinfo {pages} {1495}
  (\bibinfo {year} {2001})}\BibitemShut {NoStop}%
\bibitem [{\citenamefont {Huse}\ \emph {et~al.}(2003)\citenamefont {Huse},
  \citenamefont {Krauth}, \citenamefont {Moessner},\ and\ \citenamefont
  {Sondhi}}]{Huse2003}%
  \BibitemOpen
  \bibfield  {author} {\bibinfo {author} {\bibfnamefont {D.~A.}\ \bibnamefont
  {Huse}}, \bibinfo {author} {\bibfnamefont {W.}~\bibnamefont {Krauth}},
  \bibinfo {author} {\bibfnamefont {R.}~\bibnamefont {Moessner}}, \ and\
  \bibinfo {author} {\bibfnamefont {S.}~\bibnamefont {Sondhi}},\ }\href@noop {}
  {\bibfield  {journal} {\bibinfo  {journal} {Physical review letters}\
  }\textbf {\bibinfo {volume} {91}},\ \bibinfo {pages} {167004} (\bibinfo
  {year} {2003})}\BibitemShut {NoStop}%
\bibitem [{\citenamefont {Hermele}\ \emph {et~al.}(2004)\citenamefont
  {Hermele}, \citenamefont {Fisher},\ and\ \citenamefont
  {Balents}}]{Hermele2004}%
  \BibitemOpen
  \bibfield  {author} {\bibinfo {author} {\bibfnamefont {M.}~\bibnamefont
  {Hermele}}, \bibinfo {author} {\bibfnamefont {M.~P.}\ \bibnamefont {Fisher}},
  \ and\ \bibinfo {author} {\bibfnamefont {L.}~\bibnamefont {Balents}},\
  }\href@noop {} {\bibfield  {journal} {\bibinfo  {journal} {Physical Review
  B}\ }\textbf {\bibinfo {volume} {69}},\ \bibinfo {pages} {064404} (\bibinfo
  {year} {2004})}\BibitemShut {NoStop}%
\bibitem [{\citenamefont {Squires}(2012)}]{Squires2012}%
  \BibitemOpen
  \bibfield  {author} {\bibinfo {author} {\bibfnamefont {G.~L.}\ \bibnamefont
  {Squires}},\ }\href@noop {} {\emph {\bibinfo {title} {Introduction to the
  theory of thermal neutron scattering}}}\ (\bibinfo  {publisher} {Cambridge
  university press},\ \bibinfo {year} {2012})\BibitemShut {NoStop}%
\bibitem [{Note4()}]{Note4}%
  \BibitemOpen
  \bibinfo {note} {This recipe would imply the need of an impossible field
  $\protect \textbf {S}_m$, such that two tetrahedra sharing a vertex could
  have all their spins pointing in.}\BibitemShut {Stop}%
\bibitem [{Note5()}]{Note5}%
  \BibitemOpen
  \bibinfo {note} {For more details on this method please see D. Slobinsky and
  R. A. Borzi, in preparation (2019).}\BibitemShut {Stop}%
\bibitem [{\citenamefont {Guruciaga}\ \emph {et~al.}(2016)\citenamefont
  {Guruciaga}, \citenamefont {Tarzia}, \citenamefont {Ferreyra}, \citenamefont
  {Cugliandolo}, \citenamefont {Grigera},\ and\ \citenamefont
  {Borzi}}]{Guruciaga2016}%
  \BibitemOpen
  \bibfield  {author} {\bibinfo {author} {\bibfnamefont {P.}~\bibnamefont
  {Guruciaga}}, \bibinfo {author} {\bibfnamefont {M.}~\bibnamefont {Tarzia}},
  \bibinfo {author} {\bibfnamefont {M.}~\bibnamefont {Ferreyra}}, \bibinfo
  {author} {\bibfnamefont {L.}~\bibnamefont {Cugliandolo}}, \bibinfo {author}
  {\bibfnamefont {S.~A.}\ \bibnamefont {Grigera}}, \ and\ \bibinfo {author}
  {\bibfnamefont {R.}~\bibnamefont {Borzi}},\ }\href@noop {} {\bibfield
  {journal} {\bibinfo  {journal} {Physical Review Letters}\ }\textbf {\bibinfo
  {volume} {117}},\ \bibinfo {pages} {167203} (\bibinfo {year}
  {2016})}\BibitemShut {NoStop}%
\bibitem [{\citenamefont {Jaubert}\ \emph {et~al.}(2010)\citenamefont
  {Jaubert}, \citenamefont {Chalker}, \citenamefont {Holdsworth},\ and\
  \citenamefont {Moessner}}]{Jaubert2010Multicriticality}%
  \BibitemOpen
  \bibfield  {author} {\bibinfo {author} {\bibfnamefont {L.~D.~C.}\
  \bibnamefont {Jaubert}}, \bibinfo {author} {\bibfnamefont {J.~T.}\
  \bibnamefont {Chalker}}, \bibinfo {author} {\bibfnamefont {P.~C.~W.}\
  \bibnamefont {Holdsworth}}, \ and\ \bibinfo {author} {\bibfnamefont
  {R.}~\bibnamefont {Moessner}},\ }\href {\doibase
  10.1103/PhysRevLett.105.087201} {\bibfield  {journal} {\bibinfo  {journal}
  {Phys. Rev. Lett.}\ }\textbf {\bibinfo {volume} {105}},\ \bibinfo {pages}
  {087201} (\bibinfo {year} {2010})}\BibitemShut {NoStop}%
\bibitem [{\citenamefont {Moessner}(1998)}]{moessner1998rapcomm}%
  \BibitemOpen
  \bibfield  {author} {\bibinfo {author} {\bibfnamefont {R.}~\bibnamefont
  {Moessner}},\ }\href@noop {} {\bibfield  {journal} {\bibinfo  {journal}
  {Physical Review B}\ }\textbf {\bibinfo {volume} {57}},\ \bibinfo {pages}
  {R5587} (\bibinfo {year} {1998})}\BibitemShut {NoStop}%
\bibitem [{\citenamefont {Melko}\ and\ \citenamefont
  {Gingras}(2004)}]{Melko2004}%
  \BibitemOpen
  \bibfield  {author} {\bibinfo {author} {\bibfnamefont {R.~G.}\ \bibnamefont
  {Melko}}\ and\ \bibinfo {author} {\bibfnamefont {M.~J.}\ \bibnamefont
  {Gingras}},\ }\href@noop {} {\bibfield  {journal} {\bibinfo  {journal}
  {Journal of Physics: Condensed Matter}\ }\textbf {\bibinfo {volume} {16}},\
  \bibinfo {pages} {R1277} (\bibinfo {year} {2004})}\BibitemShut {NoStop}%
\bibitem [{\citenamefont {Jaubert}\ \emph {et~al.}(2008)\citenamefont
  {Jaubert}, \citenamefont {Chalker}, \citenamefont {Holdsworth},\ and\
  \citenamefont {Moessner}}]{Jaubert2008}%
  \BibitemOpen
  \bibfield  {author} {\bibinfo {author} {\bibfnamefont {L.~D.~C.}\
  \bibnamefont {Jaubert}}, \bibinfo {author} {\bibfnamefont {J.~T.}\
  \bibnamefont {Chalker}}, \bibinfo {author} {\bibfnamefont {P.~C.~W.}\
  \bibnamefont {Holdsworth}}, \ and\ \bibinfo {author} {\bibfnamefont
  {R.}~\bibnamefont {Moessner}},\ }\href {\doibase
  10.1103/PhysRevLett.100.067207} {\bibfield  {journal} {\bibinfo  {journal}
  {Phys. Rev. Lett.}\ }\textbf {\bibinfo {volume} {100}},\ \bibinfo {pages}
  {067207} (\bibinfo {year} {2008})}\BibitemShut {NoStop}%
\bibitem [{\citenamefont {Mito}\ \emph {et~al.}(2007)\citenamefont {Mito},
  \citenamefont {Kuwabara}, \citenamefont {Matsuhira}, \citenamefont {Deguchi},
  \citenamefont {Takagi},\ and\ \citenamefont {Hiroi}}]{Mito2007}%
  \BibitemOpen
  \bibfield  {author} {\bibinfo {author} {\bibfnamefont {M.}~\bibnamefont
  {Mito}}, \bibinfo {author} {\bibfnamefont {S.}~\bibnamefont {Kuwabara}},
  \bibinfo {author} {\bibfnamefont {K.}~\bibnamefont {Matsuhira}}, \bibinfo
  {author} {\bibfnamefont {H.}~\bibnamefont {Deguchi}}, \bibinfo {author}
  {\bibfnamefont {S.}~\bibnamefont {Takagi}}, \ and\ \bibinfo {author}
  {\bibfnamefont {Z.}~\bibnamefont {Hiroi}},\ }\href@noop {} {\bibfield
  {journal} {\bibinfo  {journal} {Journal of Magnetism and Magnetic Materials}\
  }\textbf {\bibinfo {volume} {310}},\ \bibinfo {pages} {e432} (\bibinfo {year}
  {2007})}\BibitemShut {NoStop}%
\bibitem [{\citenamefont {Steppke}\ \emph {et~al.}(2017)\citenamefont
  {Steppke}, \citenamefont {Zhao}, \citenamefont {Barber}, \citenamefont
  {Scaffidi}, \citenamefont {Jerzembeck}, \citenamefont {Rosner}, \citenamefont
  {Gibbs}, \citenamefont {Maeno}, \citenamefont {Simon}, \citenamefont
  {Mackenzie} \emph {et~al.}}]{Steppke2017}%
  \BibitemOpen
  \bibfield  {author} {\bibinfo {author} {\bibfnamefont {A.}~\bibnamefont
  {Steppke}}, \bibinfo {author} {\bibfnamefont {L.}~\bibnamefont {Zhao}},
  \bibinfo {author} {\bibfnamefont {M.~E.}\ \bibnamefont {Barber}}, \bibinfo
  {author} {\bibfnamefont {T.}~\bibnamefont {Scaffidi}}, \bibinfo {author}
  {\bibfnamefont {F.}~\bibnamefont {Jerzembeck}}, \bibinfo {author}
  {\bibfnamefont {H.}~\bibnamefont {Rosner}}, \bibinfo {author} {\bibfnamefont
  {A.~S.}\ \bibnamefont {Gibbs}}, \bibinfo {author} {\bibfnamefont
  {Y.}~\bibnamefont {Maeno}}, \bibinfo {author} {\bibfnamefont {S.~H.}\
  \bibnamefont {Simon}}, \bibinfo {author} {\bibfnamefont {A.~P.}\ \bibnamefont
  {Mackenzie}},  \emph {et~al.},\ }\href@noop {} {\bibfield  {journal}
  {\bibinfo  {journal} {Science}\ }\textbf {\bibinfo {volume} {355}},\ \bibinfo
  {pages} {9398} (\bibinfo {year} {2017})}\BibitemShut {NoStop}%
\bibitem [{Note6()}]{Note6}%
  \BibitemOpen
  \bibinfo {note} {Much bigger values can be reached depending on the progress
  of this technology~\cite {Hicks2014piezo} and the magnetoelasticity of the
  material.}\BibitemShut {Stop}%
\bibitem [{Note7()}]{Note7}%
  \BibitemOpen
  \bibinfo {note} {Materials with $T_N$ as small as $\sim 0.3 K$ have been
  synthesized \cite {Petit2016,Lefrancois2017,Singh2008}. Their physics is more
  complicated than that described by the classical Hamiltonian in Eq.~\ref
  {hamiltonian}, but they are promising options to check under the proposed
  conditions.}\BibitemShut {Stop}%
\bibitem [{Note8()}]{Note8}%
  \BibitemOpen
  \bibinfo {note} {It is easy to check that, although degenerated, the entropy
  of this phase is subextensive}\BibitemShut {NoStop}%
\bibitem [{\citenamefont {Moessner}\ and\ \citenamefont
  {Sondhi}(2003)}]{Moessner03}%
  \BibitemOpen
  \bibfield  {author} {\bibinfo {author} {\bibfnamefont {R.}~\bibnamefont
  {Moessner}}\ and\ \bibinfo {author} {\bibfnamefont {S.~L.}\ \bibnamefont
  {Sondhi}},\ }\href {\doibase 10.1103/PhysRevB.68.064411} {\bibfield
  {journal} {\bibinfo  {journal} {Phys. Rev. B}\ }\textbf {\bibinfo {volume}
  {68}},\ \bibinfo {pages} {064411} (\bibinfo {year} {2003})}\BibitemShut
  {NoStop}%
\bibitem [{\citenamefont {Nagle}\ \emph {et~al.}(1989)\citenamefont {Nagle},
  \citenamefont {Yokoi},\ and\ \citenamefont {Bhattacharjee}}]{Nagle1989}%
  \BibitemOpen
  \bibfield  {author} {\bibinfo {author} {\bibfnamefont {J.~F.}\ \bibnamefont
  {Nagle}}, \bibinfo {author} {\bibfnamefont {C.~S.}\ \bibnamefont {Yokoi}}, \
  and\ \bibinfo {author} {\bibfnamefont {S.~M.}\ \bibnamefont
  {Bhattacharjee}},\ }in\ \href@noop {} {\emph {\bibinfo {booktitle} {Phase
  transitions and critical phenomena}}},\ Vol.~\bibinfo {volume} {13},\
  \bibinfo {editor} {edited by\ \bibinfo {editor} {\bibfnamefont
  {C.}~\bibnamefont {Domb}}\ and\ \bibinfo {editor} {\bibfnamefont
  {J.}~\bibnamefont {Lebowitz}}}\ (\bibinfo  {publisher} {Academic Press New
  York},\ \bibinfo {year} {1989})\ Chap.~\bibinfo {chapter} {2}, pp.\ \bibinfo
  {pages} {235--304}\BibitemShut {NoStop}%
\bibitem [{\citenamefont {Bramwell}(2017)}]{Bramwell2017}%
  \BibitemOpen
  \bibfield  {author} {\bibinfo {author} {\bibfnamefont {S.~T.}\ \bibnamefont
  {Bramwell}},\ }\href@noop {} {\bibfield  {journal} {\bibinfo  {journal}
  {Nature communications}\ }\textbf {\bibinfo {volume} {8}},\ \bibinfo {pages}
  {2088} (\bibinfo {year} {2017})}\BibitemShut {NoStop}%
\bibitem [{\citenamefont {Hicks}\ \emph {et~al.}(2014)\citenamefont {Hicks},
  \citenamefont {Barber}, \citenamefont {Edkins}, \citenamefont {Brodsky},\
  and\ \citenamefont {Mackenzie}}]{Hicks2014piezo}%
  \BibitemOpen
  \bibfield  {author} {\bibinfo {author} {\bibfnamefont {C.~W.}\ \bibnamefont
  {Hicks}}, \bibinfo {author} {\bibfnamefont {M.~E.}\ \bibnamefont {Barber}},
  \bibinfo {author} {\bibfnamefont {S.~D.}\ \bibnamefont {Edkins}}, \bibinfo
  {author} {\bibfnamefont {D.~O.}\ \bibnamefont {Brodsky}}, \ and\ \bibinfo
  {author} {\bibfnamefont {A.~P.}\ \bibnamefont {Mackenzie}},\ }\href@noop {}
  {\bibfield  {journal} {\bibinfo  {journal} {Review of Scientific
  Instruments}\ }\textbf {\bibinfo {volume} {85}},\ \bibinfo {pages} {065003}
  (\bibinfo {year} {2014})}\BibitemShut {NoStop}%
\bibitem [{\citenamefont {Singh}\ \emph {et~al.}(2008)\citenamefont {Singh},
  \citenamefont {Saha}, \citenamefont {Dhar}, \citenamefont {Suryanarayanan},
  \citenamefont {Sood},\ and\ \citenamefont {Revcolevschi}}]{Singh2008}%
  \BibitemOpen
  \bibfield  {author} {\bibinfo {author} {\bibfnamefont {S.}~\bibnamefont
  {Singh}}, \bibinfo {author} {\bibfnamefont {S.}~\bibnamefont {Saha}},
  \bibinfo {author} {\bibfnamefont {S.}~\bibnamefont {Dhar}}, \bibinfo {author}
  {\bibfnamefont {R.}~\bibnamefont {Suryanarayanan}}, \bibinfo {author}
  {\bibfnamefont {A.}~\bibnamefont {Sood}}, \ and\ \bibinfo {author}
  {\bibfnamefont {A.}~\bibnamefont {Revcolevschi}},\ }\href@noop {} {\bibfield
  {journal} {\bibinfo  {journal} {Physical Review B}\ }\textbf {\bibinfo
  {volume} {77}},\ \bibinfo {pages} {054408} (\bibinfo {year}
  {2008})}\BibitemShut {NoStop}%
\end{thebibliography}%

\end{document}